\documentclass[a4paper,fleqn,usenatbib]{mnras}

\usepackage[T1]{fontenc}
\usepackage{ae,aecompl}
\usepackage{graphicx}
\usepackage{amsmath}
\usepackage{amssymb}

\def\d{{\rm d}}
\def\Rg{R_{\rm g}}
\def\cs{c_{\rm s}}

\def\OmK{\Omega_{\rm K}}
\def\vR{v_{R}}

\def\Qvis{Q_{\rm vis}}
\def\Qadv{Q_{\rm adv}}
\def\Qv{Q_{\nu}}
\def\Qz{Q_{\rm out}}
\def\MBH{M_{\rm BH}}
\def\Msolar{M_{\odot}}
\def\Msuns{M_{\odot}~{\rm s}^{-1}}
\def\erg{{\rm erg}~{\rm s}^{-1}}

\title[Compact binary merger: outflows and kilonova]
{Compact binary merger and kilonova: outflows from remnant disc}

\author[Yi et al.]{
Tuan Yi, Wei-Min Gu \thanks{E-mail:guwm@xmu.edu.cn},
Tong Liu, Rajiv Kumar, Hui-Jun Mu, and Cui-Ying Song
\\
Department of Astronomy, Xiamen University, Xiamen, Fujian 361005, China}

\date{Accepted XXX. Received YYY; in original form ZZZ}

\pubyear{2018}

\begin{document}
\label{firstpage}
\pagerange{\pageref{firstpage}--\pageref{lastpage}}
\maketitle

\begin{abstract}
Outflows launched from a remnant disc of compact binary merger
may have essential contribution to the kilonova emission.
Numerical calculations are conducted in this work to study the structure
of accretion flows and outflows.
By the incorporation of limited-energy advection in the hyper-accretion
discs, outflows occur naturally from accretion flows due to imbalance
between the viscous heating and the sum of the advective and radiative cooling.
Following this spirit,
we revisit the properties of the merger outflow ejecta.
Our results show that around $10^{-3} \sim 10^{-1} \Msolar$
of the disc mass can be launched as powerful outflows.
The amount of unbound mass varies with the disc mass and the viscosity.
The outflow-contributed peak luminosity is around
$10^{40} \sim 10^{41} \erg$.
Such a scenario can account for the observed kilonovae associated with
short gamma-ray bursts, including the recent event AT2017gfo (GW170817).
\end{abstract}


\section{Introduction} \label{sec:intro}

The study of the origin of short-duration gamma-ray bursts (SGRBs)
is of great importance in astrophysics.
This high energy astrophysical phenomenon,
as a subclass of gamma-ray bursts (GRBs),
is characterized by their intense flux of gamma-ray photons in only up to a few seconds
(typically less than two seconds).
Therefore it must conceal the most violent activities at its very centre.
It is commonly believed that SGRBs could be powered by
the accretion of the remnant disc after the coalescence of compact binaries
\citep[e.g.,][]{Narayan1992,Nakar2007,Berger2014},
either double neutron star (NS-NS) binaries or black hole-neutron star
(BH-NS) binaries.
A recent inspiring progress is the detection of the gravitational wave (GW)
event GW170817 from binary NS merger \citep{Abbott2017a}
and a coincident short-GRB 170817A \citep[][also observed from
the all bands of the electromagnetic radiation]{Abbott2017b,LIGO2017a}.

The merger of compact binaries with at least one NS may be accompanied with
radioactive decay of r-process elements
(rapid neutron capture nucleosynthesis),
which is known as kilonova/macronova.
\citep[e.g.,][]{Li1998,Metzger2010,Metzger2012,Jin2013,Kasen2013,Tanaka2013,Yu2013,
Gao2015,Kasen2015,Rosswog2015,Kawaguchi2016,Fernandez2017,Tanaka2016,
Chornock2017,Drout2017,Evans2017,Li2017,LIGO2017b,Metzger2017,Murguia2017b,Tanaka2017}.
The first confirmed kilonova is associated with GRB $130603$B
\citep{Tanvir2013,Berger2013,Fan2013},
which is about $1000$ times brighter than the nova
\citep[hence named `kilonova' by ][]{Metzger2010},
$10$ to $100$ times fainter than the supernova.
These events have long afterglow in optical band for several hours
to days (up to a week).
For example, the optical transient coincident with GW170817 was found
by the Swope Telescope \citep[SSS17a][]{Coulter2017},
with a total radiated energy of $1.7 \times 10^{47}$~erg over 18 days
\citep{Drout2017}.

The basic ingredients that power the kilonova
are the central compact object and the ejected materials (ejecta).
The merger can result in a stellar-mass BH surrounded by a remnant disc (torus)
or a magnetar
\citep[e.g.,][]{Dun1992,Usov1992,Dai1998,Dai2006,Metzger2011,Lv2015},
both of which can eject neutron rich materials.
For instance, the kilonova produced by GW170817
has ejected $\gtrsim 10 \%$ ($10^{-3} - 10^{-2} \Msolar$)
of the matter from the merger \citep{LIGO2017b}.
The ejecta can generally be classified into two types \citep{Fernandez2016}:
(a)~dynamical ejecta originates from the tidal tail
of the BH-NS merger or the heating by shock interface of the NS-NS merger;
(b)~outflows/wind that arises from the remnant disc.
As pointed by \citet{Metzger2017},
ejecta from outflows can rival or even dominate over the dynamical ejecta.

\begin{figure*}
  \centering
  \includegraphics[width=1.0 \linewidth]{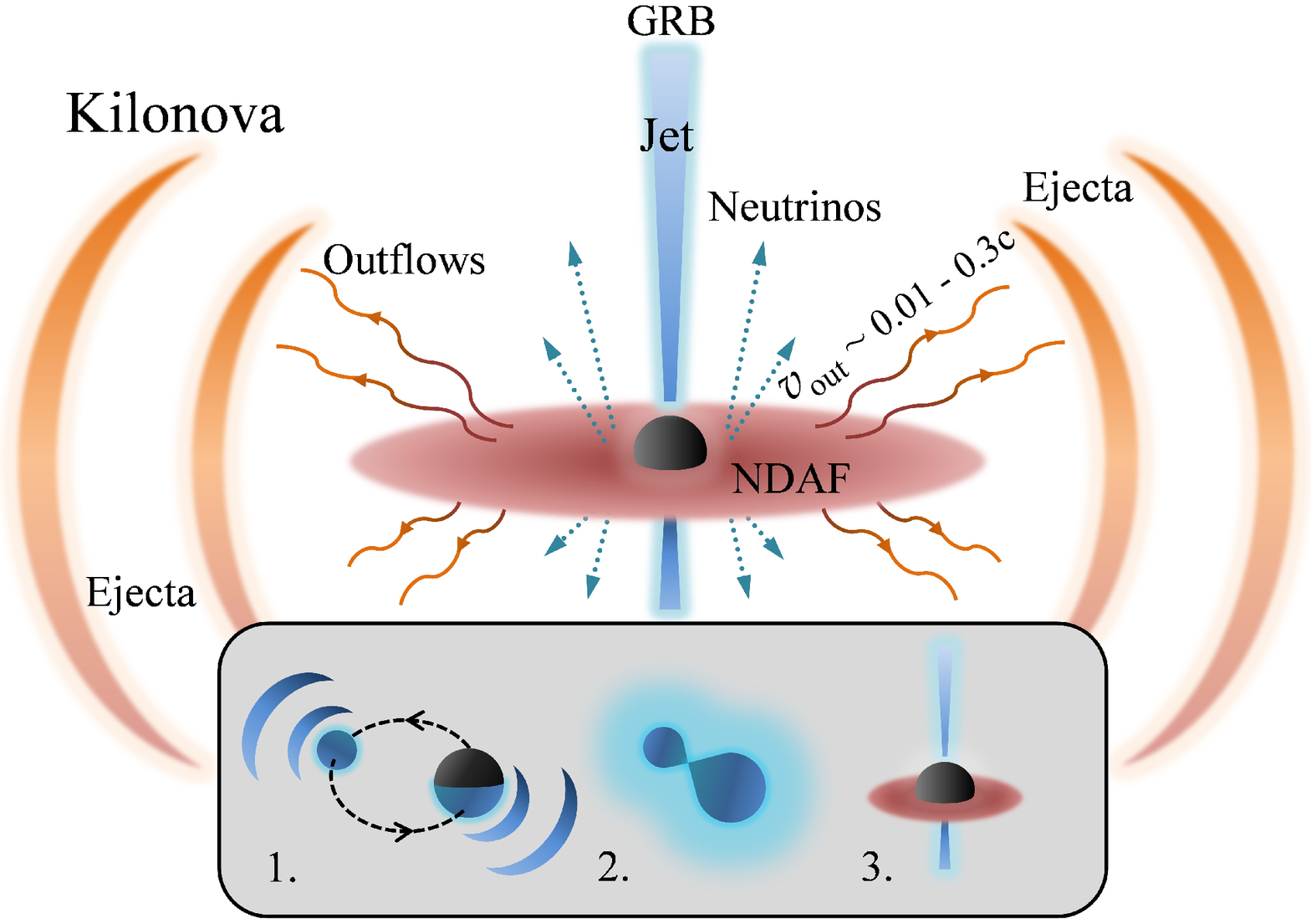}
  \caption{Illustration of a compact binary merger (NS-NS/BH-NS binary)
  with strong outflows that powers a kilonova.
  $1.$ Pre-merger stage, the binary with GW emission.
  $2.$ The coalescence of the binary.
  $3.$ The formation of a hyper-accretion disc around the BH after the merger. }
  \label{F:01}
\end{figure*}

The remnant disc launching such outflows/wind has the following properties:
(a)~extremely high accretion rates ($10^{-3}\Msuns \la \dot M \la 10\Msuns$);
(b)~extremely high density ($\rho \sim 10^{8}-10^{12} \rm g~cm^{-3}$)
and high temperature ($T \sim 10^{10}-10^{11} \rm K$).
\citet{Popham1999} studied this kind of accretion flows
and found that the neutrino emission dominates the cooling of the flow
due to the annihilation of neutrino and antineutrino pairs.
The annihilation could give rise to an energy output up to $10^{52}\erg$,
which is enough to power the observed SGRBs.
They named such flows as neutrino-dominated accretion flows (NDAFs)
(also known as the neutrino-cooled discs).
For more details on structure and properties of the NDAFs, one can refer to the works:
\citet{Popham1999,Narayan2001,Di2002,Kohri2002,Kohri2005,Gu2006,Liu2007,
Chen2007,Kawanaka2012,Kawanaka2013,Zalamea2011,Xue2013,Cao2014,Xie2016},
or a recent review \citet{Liu2017}.

Outflows exist in many accretion systems such as low-mass X-ray binaries \citep{Fender2004}
and  active galactic nuclei \citep{Terashima2001,Ganguly2008,Pounds2009}.
For example, our Galactic centre Sgr ${\rm A}^{*}$
harbors a super-massive BH (SMBH) of $3.6$ million solar masses,
and more than 99\% of the accreted mass
escape from the surrounding disc according to observations \citep{Wang2013}.
Magneto-hydrodynamical (MHD) simulations
have found that outflows exist in both optically thin flows
\citep[][]{Yuan2012a,Yuan2012b}
and optically thick flows \citep[][]{Jiang2014,Sadow2015,Sadow2016}.
Recent simulations by \citet{Jiang2017}
have found that outflows are formed with speed
$v_{\rm out} \sim 0.1 - 0.4c$ for super-Eddington accretion discs around SMBHs.
For the hyper-accretion disc,
recent three-dimensional general-relativistic MHD
(GRMHD) simulations by \citet{Siegel2017}
have shown that about $20\%$ of the initial torus mass is unbound
and form powerful outflows with speed $v_{\rm out} \sim 0.03 - 0.1c$.
The outflows can be driven by the gradient of the gas pressure,
the radiation pressure \citep{Proga2003,Ohsuga2007},
or the magnetic pressure \citep{Samadi2016}.
In addition, outflows may be driven by neutrino pressure \citep{Per2014,Murguia2017a}
at extremely high accretion rates ($\dot{M} \gtrsim 1 \Msuns$).
On the other hand, \citet{Gu2015}
proposed that for the case of low radiative cooling efficiency,
outflows should occur inevitably owing to the limited-energy advection,
no matter the flow is optically thin or thick.
A similar issue on the outflows of radiation-pressure supported accretion disc
was investigated by \citet{Gu2012}.

In this work, the central engine powering the kilonova
associated with SGRBs is revisited.
For the first time, we incorporate the limited-energy advection within NDAFs.
Consequently, the outflows can be naturally launched from the hyper-accretion disc.
In Section~\ref{sec:equa} our model is presented to describe the central engine of kilonova.
Numerical results and analyses are presented in Section~\ref{sec:resu}.
Discussion of implications and applications of our model are made in Section~\ref{sec:conc}.

\section{Theoretical model} \label{sec:equa}

Figure~\ref{F:01} is an illustration of the NS-NS/BH-NS merger and
the remnant disc formed after the binary merger.
Bipolar jet is launched by the Blandford-Znajek (BZ) process \citep[]{Blandford1977}.
In the inner part of the accretion disc,
the neutrino-antineutrino annihilation leads to a large amount of neutrino radiation.
These two mechanisms are believed to be the possible energy source of the GRBs
\citep[e.g.,][]{Liu2015,Song2015,Song2016}.
Outflows from the disc can supply the neutron rich materials for the generation of kilonova.

In the present work, the accretion disc is assumed to be steady and axisymmetric,
surrounding the merged BH.
The pseudo-Newtonian potential is adopted to describe the spacetime around the BH,
$\Psi = - G \MBH / (R-\Rg)$ \citep{PW80},
where $\MBH$ is the BH mass, and $\Rg = 2G\MBH / c^2$ is the Schwarzschild radius.

In order to examine the outflows from the disc,
the equations of continuity, motion, energy,
and the equation of state are assembled to describe the accretion flows.
The continuity equation is
\begin{equation}\label{continuity}
\dot{M} = -4 \pi \rho H R \vR \ ,
\end{equation}
where $\dot{M}$ is the accretion rate,
$\rho$ the density,
$H$ the half thickness of the disc,
and $\vR$ the radial velocity.
The equation of motion has three components.
For the radial momentum equation we assume
\begin{equation}\label{momentumr}
\Omega = \OmK = \frac{1}{R-\Rg}  \sqrt{\frac{G \MBH}{R}} \ ,
\end{equation}
with $\OmK$ being the Keplerian angular velocity.
The hydrostatic balance in the vertical direction takes as the usual simplified form:
\begin{equation}\label{momentumz}
H =\frac{\cs}{\OmK} \ ,
\end{equation}
where $\cs =\sqrt{P / \rho}$ is the isothermal sound speed,
with $P$ being the total pressure.
The azimuthal momentum can be simplified as \citep{Gu2006}
\begin{equation}\label{momentuma}
\vR = -\alpha \cs \frac{H}{R} f^{-1} g \ ,
\end{equation}
where $g = - \d \ln \OmK / \d \ln R$ and $f = 1 - j / \OmK  R^2$,
with $j$ being the specific angular momentum per unit mass accreted by the BH.
$\alpha$ is the viscous parameter.

The equation of state is \citep[e.g.,][]{Di2002}
\begin{equation}\label{pressure}
P = P_{\rm gas} + P_{\rm rad} + P_{\rm deg} + P_{\rm \nu} \ ,
\end{equation}
where the total pressure $P$ is composed of
the gas pressure $P_{\rm gas}$,
radiation pressure of photons $P_{\rm rad}$,
degenerate pressure of electrons $ P_{\rm deg}$,
and the radiation pressure of neutrinos  $P_{\rm \nu}$.
The equation of energy is expressed as
\begin{equation}\label{energy}
\Qvis = \Qadv  + \Qv  + \Qz \ .
\end{equation}
This equation shows the balance between the viscous heating $\Qvis$
and the sum of advective cooling $\Qadv$,
neutrino cooling $\Qv$, and the cooling due to outflows $\Qz$.
The expressions of $P_{\rm gas}$, $P_{\rm rad}$, $ P_{\rm deg}$, $P_{\rm \nu}$,
$\Qvis$, $\Qadv$, and $\Qv$ can be found in the previous paper \citet{Yi2017}.
One can refer to \citet{Popham1995,Di2002,Kohri2005},
for more physics of the neutrino-cooled disc described by the equations.

As mentioned in Section~\ref{sec:intro},
\citet{Gu2015} studied the accretion flows by comparing
the vertically integrated advection cooling rate with the viscous heating rate,
the former is found to be generally less than 30\% of the latter.
This hints that the outflows should rise up naturally to balance
the heating owing to the limited-energy advection and radiation.
This constraint is adopted in our analyses and calculations.
We define $f_{\rm adv} = \Qadv / \Qvis$ as the advection factor,
$f_{\rm \nu} = \Qv / \Qvis$ the neutrino cooling factor,
and $f_{\rm out} = \Qz / \Qvis$ the outflows factor.
These factors compete each other to gain the balance in the accretion flows.
The maximum value of $f_{\rm adv}$ is set to be $0.2$,
which restricts the advection portion, and therefore the excess viscous
heating energy will be dispensed to the outflows.

The standard $\alpha$-disc prescription \citep{Shakura1973} is adopted in this work,
namely the kinematic viscosity coefficient $\nu = \alpha \cs H$.
The viscous parameter $\alpha$ is subject to uncertainty \citep{Potter2014},
so the dependence of the results to $\alpha$ is studied in this work.
Since the quantities in Equations~(\ref{continuity})$\sim$(\ref{energy})
are functions of temperature $T$ and density $\rho$,
with the given parameters $\MBH$, $j$, $\dot{M}$ and $\alpha$, the profiles
of $T$ and $\rho$ can be evaluated at each point along the radius.
The mass of central BH is taken to be $\MBH = 3 M_{\odot}$,
the specific angular momentum $j = 1.83 c\Rg$
(close to the Keplerian angular momentum at the marginally stable orbit
$l_{\rm K}|_{3 \Rg} = 1.837c\Rg$),
and the accretion rate is set in the range $10^{-2}\Msuns \la \dot M \la 10\Msuns$.
Four different values of viscous parameter are examined,
i.e., $\alpha$ = 0.01, 0.02, 0.05, and 0.1.

\section{Numerical Results} \label{sec:resu}

\subsection{Outflow regions and strength}

\begin{figure}
\includegraphics[width=\columnwidth]{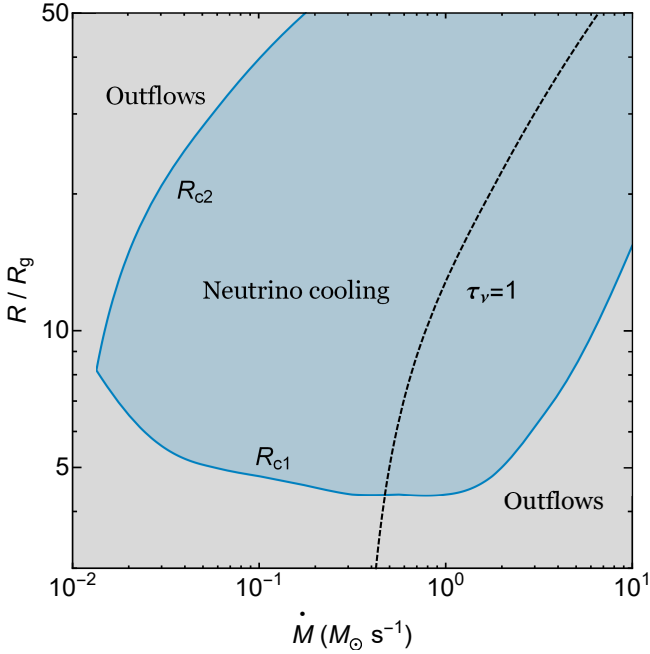}
\caption{The neutrino cooling region (sky blue) and the outflows region
(gray) for different accretion rates, where $\alpha = 0.02$ is adopted.
The black dashed line represents the photosphere for the neutrinos.}
\label{F:02}
\end{figure}

In this section, we present numerical results of the accretion flows with
outflows. Outflows were ignored in most previous works on NDAFs
\citep[e.g.,][]{Popham1999,Narayan2001,Di2002,Kohri2002,Gu2006}.
When the energy equilibrium is re-examined with limited-energy advection,
i.e., the balance of Equation~(\ref{energy}),
we can find the regions where outflows ought to occur.
As shown in Figure~\ref{F:02} where $\alpha =0.02$ is adopted
\citep{Hirose2009}, for a given mass accretion rate $\dot M$,
the flow can be divided into three regions by two critical radii
$R_{\rm c1}$ and $R_{\rm c2}$.
The inner and outer regions (gray) correspond to the cases that outflows ought to occur,
whereas the middle region (sky blue) corresponds to the cases that
the sum of neutrino cooling and advective cooling
can balance the viscous heating,
therefore outflows are not necessary.
The black dashed line represents the photosphere where the optical depth for the neutrinos equals unity.
The physics behind the figure is that,
for low accretion rates
the outer part of the disc has relatively low density and temperature.
In such case the neutrino cooling is quite inefficient.
In addition, as mentioned in Section~\ref{sec:intro},
the energy advection is also limited. Thus, the sum of the neutrino cooling
and the advective cooling cannot balance the viscous heating, and therefore
outflows will be launched and take away part of the viscous heating.
On the other hand, for high accretion rates ($\dot M \gtrsim 1 \Msuns$),
neutrino emission becomes much stronger,
but the inner part of the disc becomes optically thick to neutrinos,
so a considerable amount of neutrinos will be trapped.
Thus the efficiency of neutrino cooling
is low compared to the large viscous heating.
As a consequence, the outflows at the inner part of the disc will be enhanced.

In order to calculate the energy carried along the outflows,
the accretion discs for different masses and sizes are considered to estimate the accretion timescales.
The mass of the disc $M_{\rm disc}$ is related to the total mass and the mass ratio of the binary,
it can also be altered by the dynamical process of the merger.
\citet{Shen2017} discussed the possible value of $M_{\rm disc}$
and suggested an upper limit of $M_{\rm disc} \lesssim 0.3 \Msolar$.
Numerical simulations by \citet{Just2015b} found a similar range of $ 0.03-0.3 \Msolar$.
Typically the disc will extend itself up to a few hundred kilometers from the centre \citep{Siegel2017}.

The outer boundary of the disc radius is set in the range of
$10\Rg \leqslant R_{\rm out} \leqslant 50\Rg$,
 while the accretion rate $\dot M$ is taken from $10^{-3} - 10~\Msuns$.
Thus the masses of the disc can be calculated by the radial integration of the density.
The timescale of the accretion can then be estimated by
\begin{equation}\label{time}
\Delta t_{\rm acc} \thickapprox \int^{R_{\rm out}}_{R_{\rm in}} -\frac{1}{v_{R}}~\d R
= \frac{M_{\rm disc}}{\dot{M}} \ .
\end{equation}

\begin{figure}
\includegraphics[width=\columnwidth]{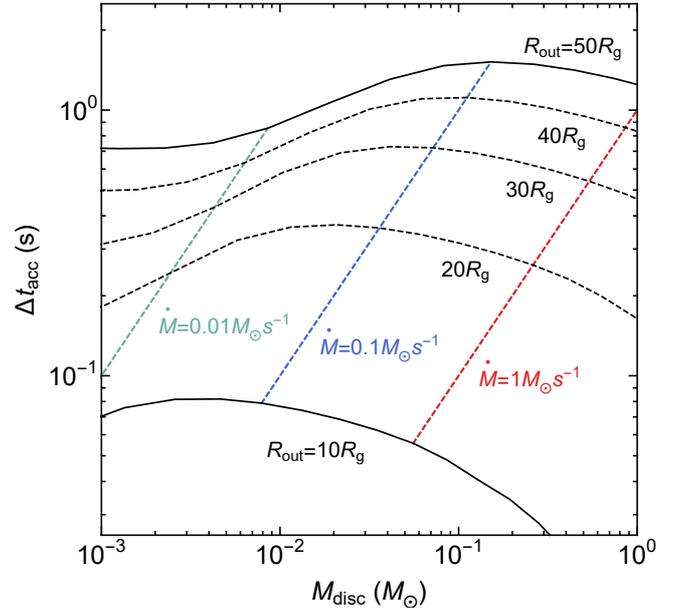}
\caption{The estimated accretion timescales for discs with different masses,
sizes and accretion rates, where $\alpha = 0.02$.}
\label{F:03}
\end{figure}

The outflows will take launch within $\Delta t_{\rm acc}$,
before the remnant disc is exhausted.
Figure~\ref{F:03} shows the variation of
$\Delta t_{\rm acc}$ with respect to $M_{\rm disc}$.
Two black solid lines mark the range of the disc radius
($10 \Rg$ and $50 \Rg$),
and black dashed lines mark three typical radii in between.
Three dashed colour lines represent the results
for $\dot M = 0.01$ (green),
$\dot M = 0.1$ (blue), and $1\Msuns$ (red).
The results show that the accretion timescales vary from
a few tens of milliseconds up to around two seconds,
which are consistent with the durations of typical SGRBs ($\lesssim$ 2s).

With known masses, sizes, accretion durations, and outflows regions of the discs,
the energy carried away by the outflows can be investigated in the first place.
The total viscous heating can be
evaluated by the radial integration of the viscous heating:
\begin{equation}\label{heating}
A_{\rm vis} = \int^{R_{\rm out}}_{R_{\rm in}}
4 \pi R \cdot Q_{\rm vis}~\d R
\approx \frac{1}{16} \dot Mc^{2}\ .
\end{equation}
The dimensionless total cooling rate factors are defined as:
$K_{\nu} = \int^{R_{\rm out}}_{R_{\rm in}}
4 \pi R \cdot Q_{\nu} \d R/A_{\rm vis}$,
$K_{\rm adv} = \int^{R_{\rm out}}_{R_{\rm in}}
4 \pi R \cdot Q_{\rm adv} \d R/A_{\rm vis}$, and
$K_{\rm out} = \int^{R_{\rm out}}_{R_{\rm in}}
4 \pi R \cdot Q_{\rm out} \d R/A_{\rm vis}$,
for neutrinos, advection, and outflows, respectively.

\begin{figure}
\includegraphics[width=\columnwidth]{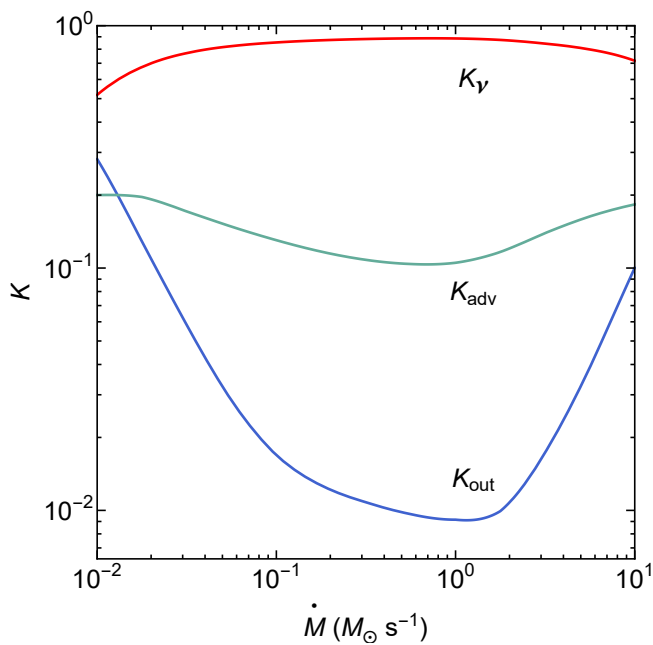}
\caption{Variations of the total outflow, advection,
and neutrino cooling factors with respect to $\dot M$,
where $R_{\rm out}=50 \Rg$ and $\alpha=0.02$.
}
\label{F:04}
\end{figure}

Figure~\ref{F:04} shows the variation of $K_{\nu}$ (red line),
$K_{\rm adv}$ (green line), and $K_{\rm out}$ (blue line) with respect to $\dot M$.
The size of the disc is taken to be $50\Rg$.
As shown in the figure,
the neutrino cooling is the highest for all given accretion rates,
the advective cooling is in between, and the outflows is the lowest.
The outflows possess a relatively large fraction ($\sim 10 \%$)
at low and high ends of the accretion rate.
The overall outflows factor is around $10^{-2} \sim 10^{-1}$.
Even at its minimum, about $1 \%$ of the viscous heating is taken by the outflows.
It is a significant amount of energy as you will see in the following analyses.

\subsection{Outflow-contributed materials and energy}

In this section,
we revisit the properties of the outflow component of the kilonova.
The goal is to estimate the mass of outflows ejecta and its contribution
to the luminosity.
The calculations are organized as follows.
Step (1): the disc outflow masses $M_{\rm out}$ are calculated.
It is derived by the kinetic energy
($E_{\rm out}=\frac{1}{2}M_{\rm out}v_{\rm out}^{2}$)
and assumed velocities ($v_{\rm out}$) of the outflows.
The dependence of the outflow mass to different disc mass $M_{\rm disc}$
and viscous parameter $\alpha$ is investigated.
Step (2): the peak luminosity $L_{\rm peak}$
of the kilonova emission is calculated
by adopting the one-zone expanding envelope model \citep{Li1998}.
The impacts of $M_{\rm out}$, $v_{\rm out}$,
and $\alpha$ to $L_{\rm peak}$ are studied.
In addition, the influence of the opacity $\kappa$ is briefly discussed.

The kinetic energy $E_{\rm out}$ is different from the total
outflows energy.
A part of the latter is required to free materials from central potential well.
Hence $E_{\rm out}$ equals the total outflows energy minus the
binding energy of the flow.
Thus the following equation holds:
\begin{equation}\label{mejec}
\frac{1}{2} M_{\rm out} v_{\rm out}^{2} =
\int^{R_{\rm out}}_{R_{\rm in}} 4 \pi R \cdot Q_{\rm out} \d R
\times \Delta t_{\rm acc} - \eta M_{\rm out} c^{2} \ ,
\end{equation}
where $Q_{\rm out}$ is computed by Equation~(\ref{energy}).
The dimensionless factor $\eta$ denotes the fraction of the rest mass energy
that is stored as the binding energy of the flow.
Equation~(\ref{mejec}) can be easily solved
for different outflows velocities $v_{\rm out}$.
The domain $0.01 \sim 0.3 c$ for $v_{\rm out}$ is investigated.
\citep{Drout2017,Fujib2017,Siegel2017}.

\begin{figure}
\includegraphics[width=\columnwidth]{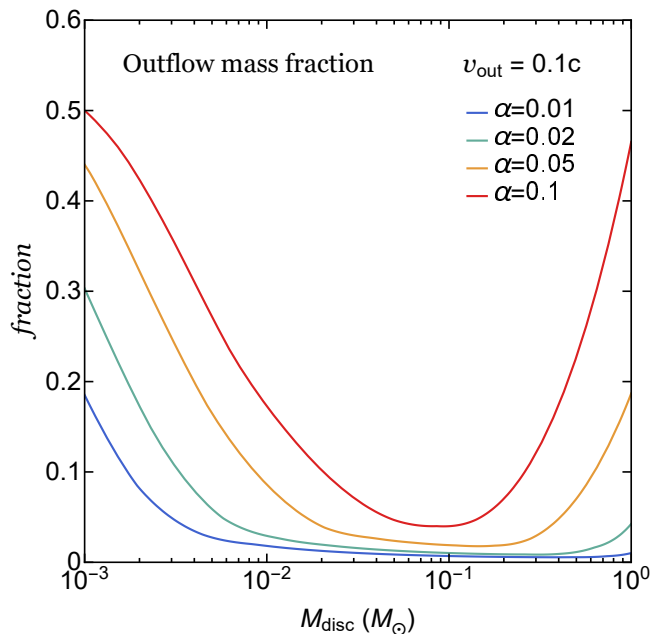}
\caption{Variations of the outflow mass fraction with the disc mass.
Colour lines represent results for $\alpha$ =0.01 (blue),
0.02 (green), 0.05 (orange), and 0.1 (red).}
\label{F:05}
\end{figure}

The total outflows energy is obtained
by integrating the $Q_{\rm out}$ from the marginally stable orbit $3 \Rg$
to the outer boundary $R_{\rm out}$.
We choose $\eta \approx 1/16$ due to the Paczy\'{n}sky-Wiita potential.
The unbound mass is found in the range $10^{-3}\sim 10^{-1} \Msolar$.
The outflow mass fraction ($M_{\rm out}/M_{\rm disc}$)
relies on $M_{\rm disc}$ and $\alpha$.
The results are shown in Figure~\ref{F:05},
with a typical outflow velocity $v_{\rm out} = 0.1c$
and a typical outer boundary $R_{\rm out} = 20 \Rg$.
For low mass remnant disc with
$M_{\rm disc} \lesssim 0.1 \Msolar$,
the fraction increases with decreasing $M_{\rm disc}$.
If $M_{\rm disc}$ goes down to $\sim 10^{-3} \Msolar$,
more than $30 \%$ of the disc materials can be launched as the form of outflows.
But the fraction heavily depends on the viscosity $\alpha$.
The influence of $\alpha$ is indicated by the colour lines.
If $\alpha$ is larger, more materials tend to be pushed away.
For larger remnant discs with
$M_{\rm disc} \gtrsim 0.1 \Msolar$,
the outflows possess a relatively small mass fraction $\lesssim 10\%$.
The viscosity has relatively mild effects on the flow.
The dependence of the results to $\alpha$
can be understood in the following way.
According to Equation~(\ref{momentuma}), large $\alpha$ means large radial
velocity. As a consequence, the ratio of the advective cooling to the neutrino
cooling becomes large. However, due to the limited-energy advection, a
large amount of energy stored in the accretion flow ought to be released
through outflows. Thus, for larger viscous parameter, the fraction of energy
owing to outflows is also larger, and therefore the outflow mass fraction
becomes larger.

The amount of merger outflows has been investigated by many simulations
\citep[e.g.][]{Fernandez2013,Just2015a,Fujib2017,Shiba2017,Siegel2017}.
Despite of different parameters were assumed among those simulations,
the outflow mass fraction was found in the range $5\% \sim 30\%$.
As shown in Figure~\ref{F:05}, our results are roughly in agreement with
the simulation results. For instance, in the GRMHD simulations of
\citet{Siegel2017}, a torus of $0.02 \Msolar$ will lose $16 \%$ of
the materials through outflows, and our calculations found that the value
is $\sim 10\%$ in the similar circumstances.
The quantitative difference between our results and the simulations
may be related to the different outer boundaries and outflow velocities.
Moreover, our results are based on a fixed upper limit of energy advection
$f_{\rm adv}^{\rm max} = 0.2$. However, the previous analyses \citep{Gu2015}
indicate a flexible upper limit, i.e. $f_{\rm adv}^{\rm max} \la 0.3$,
which can have quantitative influence on the results.
In addition, the magnetic fields are not taken into
consideration in the present work, which may also have effects on our results.

\begin{figure}
\includegraphics[width=\columnwidth]{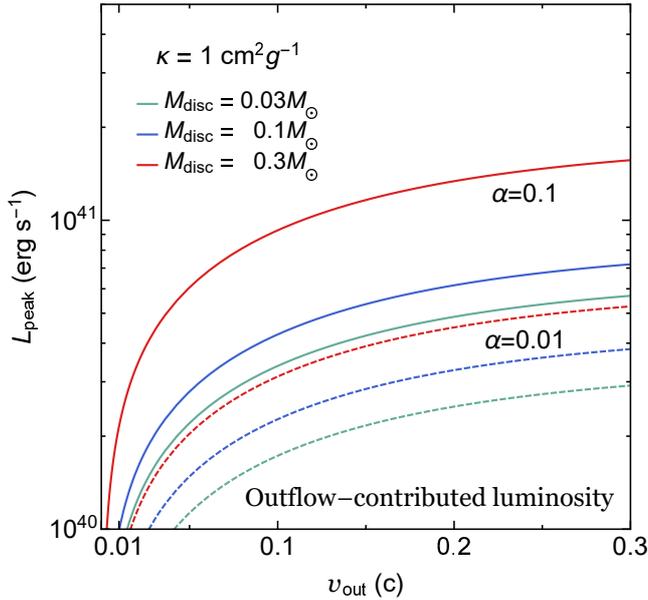}
\caption{Variation of the peak luminosity $L_{\rm peak}$
with the outflows velocity $v_{\rm out}$.
Three typical disc masses are examined, i.e.,
$M_{\rm disc}$ = 0.03$\Msolar$ (green), 0.1$\Msolar$ (blue),
and 0.3$\Msolar$ (red).
Solid and dashed lines distinguish between $\alpha = 0.1$ and $0.01$.}
\label{F:06}
\end{figure}

The kilonova ejecta can be simplified as spherically,
uniformly expanding layers of neutron rich materials.
\citep[][]{Li1998,Metzger2010,Kasen2013,Tanaka2013,
Rosswog2015,Tanaka2016,Fernandez2016,Metzger2017}.
The peak luminosity of the radioactive decay is given by
\citep{Tanaka2016}:
\begin{equation}\label{lpeak}
\begin{split}
L_{\rm peak} = 1.3 \times 10^{40} \erg
\times (\frac{\epsilon_{\rm dep}}{0.5})^{1/2}
({\frac{M_{\rm ejec}}{0.01 \Msolar}})^{0.35} \\
\times ({\frac{v}{0.1 c}})^{0.65}
({\frac{\kappa}{10~{\rm cm^{2}~g^{-1}}}})^{-0.65} \ ,
\end{split}
\end{equation}
where $\epsilon_{\rm dep} < 1$ is the fraction of energy deposition.
A typical value 0.5 for $\epsilon_{\rm dep}$ is taken.
$M_{\rm ejec}$ is specified as the outflows ejecta mass
$M_{\rm out}$ in the calculation.
$\kappa$ is the opacity of the outflows. The value of $\kappa$ is uncertain
due to the complexity of the outflow composition \citep{Barnes2013}.

Figure~\ref{F:06} shows the variations of
$L_{\rm peak}$ with respect to $v_{\rm out}$.
Three typical disc masses are examined, i.e.,
$M_{\rm disc}$ = 0.03 $\Msolar$ (green), 0.1 $\Msolar$ (blue), and 0.3 $\Msolar$ (red).
Solid and dashed lines distinguish between $\alpha = 0.1$ and $0.01$.
The opacity $\kappa$ is taken as $1~{\rm cm^{2}~g^{-1}}$.
The peak luminosity is found to be around
$10^{40} \sim 10^{41} \erg$.
Taking a typical case for example,
where the parameters are $M_{\rm disc} = 0.3 \Msolar$,
$\alpha = 0.1$, and $v_{\rm out} = 0.1c$, we derive an outflow mass
$M_{\rm out} \approx 0.037 \Msolar$.
In this case, a peak luminosity $L_{\rm peak} \approx 10^{41} \erg$ is obtained,
which is in agreement with the three kilonovae associated with GRBs
050709, 060614 \citep{Jin2015,Jin2016}, and 130603B.

As mentioned above,
the ejecta opacity can influence the luminosity significantly.
The kilonova AT2017gfo (GW170817) has a peak bolometric luminosity
$L_{\rm bol} \approx 5\times10^{41} \erg$ \citep{Cowper2017}.
In order to explain the peak luminosity of AT2017gfo,
higher opacity with $\kappa =1 \sim 10~{\rm cm^{2}~g^{-1}}$ may be ruled out.
Instead, larger $M_{\rm out}$, $v_{\rm out}$,
and lower $\kappa$ ($\lesssim 0.5~{\rm cm^{2}~g^{-1}}$)
are favoured \citep{Smartt2017}.

\section{Conclusions and discussion} \label{sec:conc}

In this work, we have extended the limited-energy advection upon NDAFs
for the first time, which naturally gives an output of energy.
By using this model along with a couple of simplified assumptions,
the outflows that contribute to kilonova have been examined in a tidy way.
In addition, the energy output and the masses of the ejecta have been
evaluated.
Our results have shown that about $10^{-3}\sim 10^{-1} \Msolar$
of the disc mass is launched to the form of outflows,
providing the materials and the energy source for the kilonova ejecta.
The peak luminosity is estimated to be around
$10^{40} \sim 10^{41} \erg$,
from the contribution by outflows.

The outflows of the hyper-accretion discs described by Equation~(\ref{energy})
are in fact taken from the excess of the advection,
which is restrained to less than $20\%$ of the total heating energy.
But this value is actually quiet uncertain,
the possible values can wander from a few percent up to around thirty percent,
due to the variation of the flow density.
However, there are strong evidence for the outflows as mentioned in Section~\ref{sec:intro},
so the issue may be the least of our concern.

\citet{Gao2017} presented the merger-nova events from \textit{Swift} SGRBs sample.
They have found three candidates GRB $050724$, $070714$B and $061006$.
The optical peak luminosity of these sources is around $10^{42} - 10^{43} \erg$,
which is a few tens to hundreds times brighter than the normal kilonova.
So the central engine of these candidates must be something different
rather than a BH surrounded by a disc.
The merger of an NS binary may result in a massive millisecond magnetar,
which can continuously provide additional source of energy ejection to power the merger ejecta
\citep{Yu2013,Siegel2016}.
An alternative way to explain such high luminosity is by considering
the vertical advection effects caused by the magnetic buoyancy
\citep{Jiang2014}.
Optically thick accretion flows are capable of trapping a significant amount of photons.
Those photons together with the in-falling matter,
will be devoured by the BH before they can even escape.
The vertical advection effect rescues part of these trapped photons,
by bringing out them with the bulk motion of the hot flows
\citep{Yi2017}.
The speed of the vertical advection is roughly the local sound speed in general,
much faster than the diffusive speed of the photons.
The existence of the outflows is a natural bridge for such energy transportation.
An extra part of the radiation can be unleashed through the out-traveling materials,
so the contribution from it will enhance the luminosity to the kilonova.

\section*{Acknowledgements}

We thank Da-Bin Lin, Shu-Jin Hou, and Mou-Yuan Sun for beneficial discussion,
and thank the referee for helpful suggestions that improved the manuscript.
This work was supported by the National Basic Research Program of China
(973 Program) under grants 2014CB845800,
the National Natural Science Foundation of China under grants 11573023,
11473022, and 11333004,
and the CAS Open Research Program of Key Laboratory for the Structure and
Evolution of Celestial Objects under grant OP201503.

\bsp
\label{lastpage}
\end{document}